\documentclass{emulateapj}

\usepackage{graphicx}
\usepackage{verbatim} % for comments

\newcommand{\vtbold}[1]{#1}               % for arXiv

\slugcomment{Accepted for publication in ApJ Letters}
\shorttitle{HIPPARCOS CALIBRATION OF THE TRGB}
\shortauthors{TABUR, KISS, \& BEDDING}

\begin{document}

\title{Hipparcos calibration of the tip of the Red Giant Branch}
\author{Vello Tabur, L\'aszl\'o~L. Kiss and
  Timothy R. Bedding}
\affil{Sydney Institute for Astronomy (SIfA), School of Physics, The University of Sydney, NSW 2006, Australia}
\email{tabur@physics.usyd.edu.au}

\label{firstpage}
\begin{abstract}

We have detected the tip of the Red Giant Branch (TRGB) in the solar
neighborhood using near infrared photometry from the 2MASS and DIRBE
catalogs, and revised Hipparcos parallaxes.  We confirm that the revised
Hipparcos parallaxes are superior to the original ones, and that this
improvement is necessary to detect the TRGB\@. We find a tip absolute
magnitude of $M_K=-6.85\pm0.03$, in agreement with that expected from
previous tip measurements of the Large Magellanic Cloud, Small Magellanic
Cloud, and Bulge. This represents the first geometric calibration of the
TRGB and extends previous calibrations, based on metal-poor globular
clusters, to solar metallicities.  We attempted to use the TRGB to confirm
the presence of the Lutz-Kelker bias, with inconclusive results.  Attempts
to detect the tip in the $I$-band also produced inconsistent results, due
to a lack of precise, homogeneous photometry for these bright stars.

\end{abstract}


%%%%%%%%%%%%%%%%%%%%%%%%%%%%%%%%%%%%%
\section{Introduction}
%%%%%%%%%%%%%%%%%%%%%%%%%%%%%%%%%%%%%

The tip of the Red Giant Branch represents the maximum absolute luminosity
achieved by first-ascent red giants, and marks the onset of helium fusion
in the degenerate cores of these
low-mass stars. Theoretical predictions, confirmed by observational evidence, indicate that the
TRGB is an excellent distance indicator, since the absolute bolometric magnitude of the tip
varies by only $\sim 0.1$ mag for a wide range of metallicities and ages
\citep{b_ibe83,b_dac90,b_sal97}. The $I$-band magnitude of the TRGB has become well-established
as a distance indicator for nearby galaxies with well-resolved Population II stars in their
halos \citep[for example, see][]{b_lee93,b_sak96,b_mou08}.

The TRGB complements other distance indicators, such as Cepheids and RR Lyrae stars, being
comparable in precision \citep{b_lee93}. Indeed, empirical evidence and computer simulations
show that space-based observations provide distances out to $\sim$12\,Mpc with a precision of
10\%, limited primarily by integration time \citep{b_mad95}. However, the TRGB is a tertiary
distance indicator, calibrated using RR Lyrae variables in Galactic globular clusters, which are
themselves secondary distance indicators.
Moreover, there is a discrepancy of $\sim$0.1mag between the TRGB and Cepheid distance scales
\citep{b_tam08}, and the metallicity dependence of Cepheid P-L relation has itself been
calibrated using the TRGB method, leading to a circular dependency \citep{b_riz07}. Thus, it
would be of considerable importance to have a direct calibration of the tip magnitude.

Although the TRGB is more easily observed in near infrared (NIR) passbands, where interstellar
reddening is reduced, both the color and luminosity of the NIR TRGB are more sensitive to
metallicity than in the $I$-band \citep{b_riz07}. Nevertheless, recent studies of the Magellanic
Clouds and Galactic Bulge have successfully identified the TRGB discontinuity in the $K$-band,
using 2MASS photometry \citep{b_cio,b_kis04,b_sch}. Moreover, the discontinuity is clearly
visible in the $K$-band P-L plot of M giants in the local solar neighborhood \citep{b_tab09a},
which prompted us to investigate further. In this Letter we present our findings, including the
first geometric calibration of the TRGB absolute magnitude.

%%%%%%%%%%%%%%%%%%%%%%%%%%%%%%%%%%%%%%%%%%%%%%%%%%%%%%%%%
\section{Data}
\label{s_samp}
%%%%%%%%%%%%%%%%%%%%%%%%%%%%%%%%%%%%%%%%%%%%%%%%%%%%%%%%%

We selected all stars with relative parallax uncertainties less than 25\%
in the revised Hipparcos catalog \citep{b_van07}\footnote{We used the
version published on VizieR on 2008 September 15, which corrected an error
that affected earlier versions of the revised catalog.  Stars with
negative parallaxes were ignored. }.
We obtained $JHK$ magnitudes and uncertainties  from
the 2MASS catalog \citep{b_cut}, using a search radius of
5\,arcsec. Being luminous and nearby, some of the stars were saturated
despite being observed with the shortest integration time (51\,ms), and were
measured using the wings of their radial profiles, resulting in large
photometric uncertainties (0.2--0.4\,mag). Additional scatter was
contributed by the use of single-epoch measurements, since all M giants are
intrinsically variable \citep{b_eye08}\vtbold{, with typical $K$-band peak-to-peak amplitudes between 0.1 and 0.25 mag \citep{b_smi}}.

Seeking a higher-precision source of NIR photometry, we extracted flux
measurements from the DIRBE catalog, which contains photometry for nearly
12000 objects sampled over 10 months \citep{b_smi}.  Following
\citet{b_whi}, we adopted a value of 630\,Jy for $m_K=0$, and used 1570\,Jy
for $m_J=0$, both of which are consistent with \citet{b_bes}. Although
DIRBE was less prone to saturation than 2MASS, its large beam-width led to
confusion between nearby sources. Using the same methodology as described
by \citet{b_tab09a}, we only selected unconfused sources (confusion flags 1
and 2 not set).  This permitted an accurate determination of mean $K$
magnitudes that are demonstrably superior to 2MASS for about half the red
giants.

Interstellar extinction is expected to be relatively small for all stars in
the sample, particularly in the $K$-band. We calculated the visual
extinctions using the interstellar extinction model of \citet{b_dri03}.
%, which showed generally small extinctions.
Extinctions for other bands were scaled using
the factors from \citet[][]{b_rie}. The greatest extinction was $A_K \sim 0.15$ mag, with
$A_K < 0.05$ mag for 98\% of the sample.

We calculated absolute magnitudes using the relation $M = m + 5 + 5 \log
\pi$, where $m$ is the extinction-corrected apparent magnitude, and $\pi$
is the geometric parallax in arcseconds.  Photometric uncertainties were
obtained directly from the 2MASS catalog, or derived from flux densities
from DIRBE. Parallax uncertainties were estimated with a first-order
approximation of $5 \log {\rm e} \; \sigma_{\pi} / \pi \approx 2.17 \;
\sigma_{\pi} / \pi$, with the total uncertainty in $M$ calculated as the
quadrature sum of photometric and parallax uncertainties\vtbold{\footnote{The data used in this analysis are available electronically from CDS, Strasbourg.}}.

\begin{figure}
 \includegraphics[scale=1.0, angle=0]{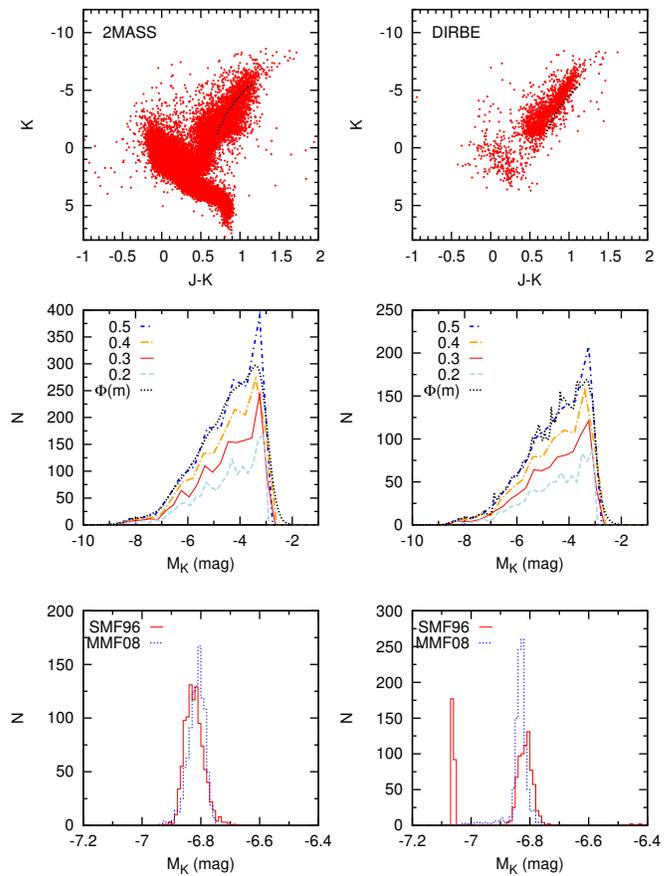}
 \caption{Top: Color-magnitude diagrams using photometry from 2MASS (left)
 and DIRBE (right) for all stars with revised Hipparcos parallaxes
 satisfying $\sigma_{\pi}/\pi \le 0.125$.  Middle: luminosity functions for
 red giants, with bin sizes as shown.  Bottom: results of applying two
 edge-detection methods to the LFs (see text).}
 \label{fig001}
\end{figure}

The color-magnitude diagrams (CMDs) are shown in the top row of Figure
\ref{fig001}, and the RGB is clearly visible.  We indicate its expected
position with a thin black line, by assuming $[{\rm Fe/H}]=0$ and adopting
the calibration of \citet{b_val04}, which is based on Galactic globular
clusters within the metallicity range $-2.12 \le {\rm [Fe/H]} \le
-0.49$. The position agrees well for our 2MASS sample, and less so for
DIRBE, although the difference is irrelevant for measuring the TRGB\@.  A
few stars around $J-K \sim 1.5$ and $M_K \sim 0$ are probably reddened by
dust. The DIRBE photometry is sparser, particularly at the faint end,
although the evolved stars near the tip appear well represented.

To be sure of identifying all stars near the tip, which contains relatively
few stars because it is a short-lived late evolutionary stage, we ignored
cataloged spectral types. Instead, we used the color-magnitude diagram
to select all stars in the range $0.5 \le
J-K \le 2.0$ and $M_K < -3$, in order to include early M giants, and to
exclude very red, dust-enshrouded stars.

%%%%%%%%%%%%%%%%%%%%%%%%%%%%%%%%%%%%%%%%%%%
%\section{Data analysis}
\section{Edge detection}
\label{s_anal}
%%%%%%%%%%%%%%%%%%%%%%%%%%%%%%%%%%%%%%%%%%%

% 089/fig1.plt

We have used two model-independent methods for edge detection (ED) to
locate the TRGB discontinuity.  Firstly, we followed \citet[][hereafter
MMF08]{b_mag08} by using an ED filter that uses a combination of small bins
for resolution and larger bins for smoothing, with a logarithmic edge
detector to account for the expected power-law distribution at the bright
end of the RGB \citep{b_men02}. Additionally, we iteratively calculated the
peak filter-response 85 times, using a range of starting offsets and bin
sizes to smooth-out sensitivity to these parameters, and subsequently used
the mean (after 3-sigma rejection) as the TRGB location. This method proved
fairly robust in the presence of noise. We limited the search for the
discontinuity to the range $-8 \le M_K \le -6$, to eliminate false
positives.

Our second method accounts for photometric and parallax uncertainties by
using a sum of Gaussians to create continuous probability function to model
the LF:
\begin{equation}\label{eq:1}
\Phi(m) = \sum_{i=1}^N {\frac{1}{\sigma_i \sqrt{2\pi}} \exp\left[
    -\frac{(m_i - m)^2}{2\sigma_i^2} \right] },
\end{equation}
where $m_i$ and $\sigma_i$ are the magnitudes and their uncertainties,
respectively, and $N$ is the total number of stars in the sample
\citep[][hereafter SMF96]{b_sak96}.  We found that a logarithmic ED filter
produced poor results and, following SMF96, adopted a filter of the form
$E(m) = \Phi(m+\Delta m) - \Phi(m-\Delta m)$, where $E(m)$ is the filter
response at magnitude $m$, and $\Delta m$ is the bin size. Previous studies
have defined the bin size as a function of photometric
uncertainty. However, unlike the RGBs in nearby galaxies, our sample is
drawn from the local solar neighborhood  and is less affected by
completeness errors or increased photometric errors toward fainter
magnitudes. Indeed, the distribution of mean errors over the range $-8 \le
M_K \le -6$ is nearly constant, leaving us free to choose a bin size on the
basis of its smoothing performance alone. We adopted a value of 0.2\,mag.

We estimated uncertainties using the bootstrapping method described by \citet{b_bab96}, which
is widely used \citep[see, for example,][]{b_men02}. We selected 1000 random samples,
each consisting of 80\% of the stars from the original sample, and determined the location of
the TRGB for each, adopting the rms scatter as our uncertainty.

%%%%%%%%%%%%%%%%%%%%%%%%%%%%%%%%%%%%
\section{Results and Discussion}
\label{s_result}
%%%%%%%%%%%%%%%%%%%%%%%%%%%%%%%%%%%%

%%%%%%%%%%%%%%%%%%%%%%%%%%%%%%%%%%%%%%%%%
\subsection{The K-band TRGB}
%%%%%%%%%%%%%%%%%%%%%%%%%%%%%%%%%%%%%%%%%%

Since samples containing stars with relatively large $\sigma_{\pi}/\pi$ tended to blur the
RGB/AGB boundary, we calculated the TRGB location for several samples selected with upper bounds
in the range $0.10 \le \sigma_{\pi}/\pi \le 0.25$.  Luminosity
functions (LFs) and bootstrapping distributions for the stars with $\sigma_{\pi} / \pi \le
0.125$ are shown in Figure~\ref{fig001}.
LFs are shown for bin sizes of 0.2--0.5 mag, together with the weighted continuous function (Eq.
\ref{eq:1}). The two LFs exhibit similar morphologies and, despite the DIRBE LF containing half
as many stars as 2MASS, both show the AGB stars above the tip, at $-9 \le M_K \le -7$.

The bottom panels show the TRGB position determined from 1000 bootstrap operations for each ED
filter, binned in 0.01 mag increments. The 2MASS photometry yields an unambiguous detection,
with nearly identical results for the two filters. The sparser DIRBE photometry produced very
similar results, although the SMF96 filter detected another peak at $M_K \sim -7.1$ which was
not real. Its significance reduced as more stars were added (with looser parallax restrictions),
simultaneously increasing the size of the peak at $M_K \sim -6.8$, implying that the SMF96
filter is more sensitive to noise. We ignored the second peak in our calculations.

Table~\ref{T1} gives the full results, listing the maximum $\sigma_{\pi}/
\pi$ used for each sample, the number of stars in the 1-mag bin below the
tip (2MASS), the tip locations for both the 2MASS and DIRBE samples using
the SMF96 and MMF08 filters.  The results for all four combinations agree
within the uncertainties.

% 2M from a04 and F:\HIPP\2007\GETMG2\SAVE.a04\PRO\PEAK2{m|s}.dat, respectively (DIRBE from peakd{m|s}.dat)
\begin{table}
  \caption{$K$-band TRGB magnitudes from 2MASS and DIRBE.}
  \setlength{\tabcolsep}{0.95mm}
  \label{T1}
{\scriptsize
  \begin{tabular}{@{}llllllll}
  \hline
%                                                                              ign.a04
  Max                  & $N_{2M}$  & $M_K^{\rm SMF,2M}$ & $M_K^{\rm MMF,2M}$ & $M_K^{\rm SMF,D}$ & $M_K^{\rm MMF,D}$ \\
  $\sigma_{\pi}/ \pi$  &      & (mag)              & (mag)              & (mag)             & (mag)             \\
  \hline
  0.100                &  123       & $-6.83\pm0.03$ & $-$6.79 $\pm$ 0.02 & $-$6.82 $\pm$ 0.03 & $-$6.84 $\pm$ 0.04 \\
  0.125                &  166       & $-6.83\pm0.03$ & $-$6.82 $\pm$ 0.03 & $-$6.82 $\pm$ 0.03 & $-$6.84 $\pm$ 0.04 \\
  0.150                &  212       & $-6.85\pm0.05$ & $-$6.87 $\pm$ 0.03 & $-$6.82 $\pm$ 0.03 & $-$6.90 $\pm$ 0.09 \\
  0.175                &  253       & $-6.85\pm0.03$ & $-$6.85 $\pm$ 0.03 & $-$6.82 $\pm$ 0.03 & $-$6.85 $\pm$ 0.06 \\
  0.200                &  295       & $-6.86\pm0.03$ & $-$6.83 $\pm$ 0.03 & $-$6.82 $\pm$ 0.03 & $-$6.85 $\pm$ 0.06 \\
  0.225                &  318       & $-6.86\pm0.03$ & $-$6.82 $\pm$ 0.03 & $-$6.82 $\pm$ 0.03 & $-$6.86 $\pm$ 0.06 \\
  0.250                &  353       & $-6.87\pm0.03$ & $-$6.85 $\pm$ 0.05 & $-$6.82 $\pm$ 0.02 & $-$6.94 $\pm$ 0.10 \\
  \hline
\end{tabular}
}
\end{table}

Early studies of the statistical robustness of the TRGB method concluded that 50--100 stars are required in the 1-mag bin below the TRGB, although more recent work suggests  that 400--500 stars are required for an accurate determination \citep{b_mad09}. Although our smallest 2MASS sample includes only 123 stars in the 1-mag bin below the tip, the largest sample (with 353 stars) yields very similar results. There is no indication of the 0.6 mag systematic shift toward fainter tip magnitudes for samples with less than 300 stars, that was found by \citet{b_mad09}. Moreover, rerunning the ED calculations by bootstrapping with 66\% and 33\% of the original sample size showed little variation in the TRGB location (but with larger uncertainties).

\citet{b_cio} measured the TRGB in the LMC and SMC, finding dereddened
values of $M_K = -6.61$ and $-$6.41, respectively. The difference of 0.2
mag is caused by the different metal abundances, which we adopt as
$Z_{\sun}=0.016$, $Z_{\rm LMC}=0.008$, and $Z_{\rm SMC}=0.004$
\citep{b_vas}. In order to quantify the TRGB's metallicity dependence,
\citet{b_fer} measured the location of the TRGB in 10 galactic globular
clusters, fitting the relation $M_K^{\rm TRGB} = -(0.59 \pm 0.11) {\rm
[Fe/H]}_{\rm CG97} - (6.97 \pm 0.15)$, which predicts $\Delta M_K^{\rm
TRGB} \sim -0.18$ mag for an increase in metallicity by a factor of
2. Using the global metallicity scale, which incorporates $\alpha$-element
abundances, \citet{b_fer06} found $M_K^{\rm TRGB} = -6.92 - 0.62$[M/H] for
metallicities to $\rm{[Fe/H]} \le +0.4$ dex.  This leads to a tip magnitude in
the range $-6.8$ to $-6.9$ for solar-like metallicity, in excellent
agreement with our result.

\citet{b_sch} examined the $K$-band luminosity function of MACHO semiregular variables in the Galactic bulge, finding a steep drop-off in number near $M_K \sim -6.8$, which they interpreted as the location of the TRGB.

\begin{table}
  \caption{$K$-band TRGB magnitudes corrected for Lutz-Kelker bias.}
  \setlength{\tabcolsep}{1.75mm}
  \label{T2}
{\scriptsize
  \begin{tabular}{@{}lllll}
  \hline
%                                                                 ing.a05
%                        05\peak2s.dat       05\peak2m.dat        05\peakds.dat        05\peakdm.dat
  Max                 & $M_K^{\rm SMF,2M}$ & $M_K^{\rm MMF,2M}$ & $M_K^{\rm SMF,D}$  & $M_K^{\rm MMF,D}$  \\
  $\sigma_{\pi}/ \pi$ & (mag)              & (mag)              & (mag)              & (mag)              \\
  \hline
  0.100               & $-$6.86 $\pm$ 0.03 & $-$6.81 $\pm$ 0.02 & $-$6.84 $\pm$ 0.03 & $-$6.86 $\pm$ 0.02 \\
  0.125               & $-$6.86 $\pm$ 0.03 & $-$6.83 $\pm$ 0.03 & $-$6.84 $\pm$ 0.03 & $-$6.86 $\pm$ 0.03 \\
  0.150               & $-$6.87 $\pm$ 0.03 & $-$6.93 $\pm$ 0.05 & $-$6.83 $\pm$ 0.04 & $-$6.86 $\pm$ 0.05 \\
  0.175               & $-$6.86 $\pm$ 0.03 & $-$6.84 $\pm$ 0.04 & $-$6.83 $\pm$ 0.03 & $-$6.82 $\pm$ 0.03 \\
  0.200               & $-$6.87 $\pm$ 0.03 & $-$6.85 $\pm$ 0.04 & $-$6.83 $\pm$ 0.03 & $-$6.83 $\pm$ 0.02 \\
  0.225               & $-$6.87 $\pm$ 0.02 & $-$6.84 $\pm$ 0.03 & $-$6.83 $\pm$ 0.03 & $-$6.83 $\pm$ 0.02 \\
  0.250               & $-$6.89 $\pm$ 0.02 & $-$6.83 $\pm$ 0.03 & $-$6.83 $\pm$ 0.03 & $-$6.81 $\pm$ 0.02 \\
  \hline
\end{tabular}
}
\end{table}

%%%%%%%%%%%%%%%%%%%%%%%%%%%%%%%%%%%%%%%%%
\subsection{Lutz-Kelker bias}
%%%%%%%%%%%%%%%%%%%%%%%%%%%%%%%%%%%%%%%%%%

We expect a sample limited by relative parallax uncertainties to be
affected by a systematic bias \citep{b_lut}, resulting in an underestimate
of stellar luminosity. Using the TRGB as a marker, we searched for evidence
of a systematic shift in its position in samples selected with
progressively larger relative parallax uncertainties.

First, we selected non-overlapping samples in blocks of 10\% relative
parallax uncertainty, but this yielded too few stars for reliable TRGB
detection. Increasing to 15\% increments did not help, so finally we
simply examined our previous samples for a systematic shift, and compared
them to a reprocessed set, where individual bias corrections had been
applied to each star using the relation $LK = -8.09 (\sigma_{\pi} / \pi)^2$
\citep{b_whi}. Table~\ref{T2} lists the results using the same nomenclature
as Table~\ref{T1}, and shows that the mean corrected tip positions are
indistinguishable from the uncorrected values, to within the errors. We
note that the corrected tip magnitudes for 2MASS are 0.02 mag brighter on
average, but the DIRBE MMF08 result is slightly fainter.

The median $\sigma_{\pi}/\pi$ for each sample is about half the
maximum value, implying an expected bias-induced shift of $\sim$0.1\,mag
between the first and last samples (10\% and 25\%).  However, the uncorrected
results (Table~\ref{T1}) show far smaller differences, and in the
wrong sense (brighter tip magnitudes when an under-estimate is
expected). Thus, we see no evidence of a systematic shift toward
fainter tip magnitudes as samples include stars with progressively larger
relative parallax uncertainties, and conclude that there are too few stars
with sufficiently precise parallaxes and/or apparent magnitudes for a
convincing detection of the Lutz-Kelker bias.

%%%%%%%%%%%%%%%%%%%%%%%%%%%%%%%%%%%%%%%%%
\subsection{Comparison of original and revised parallaxes}
%%%%%%%%%%%%%%%%%%%%%%%%%%%%%%%%%%%%%%%%%%

\begin{figure}
 \includegraphics[scale=1.0, angle=0]{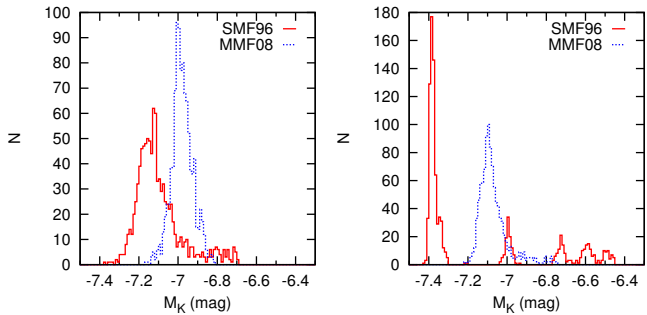}
 \caption{Same as the bottom row of Fig.~\ref{fig001} but with $M_K$
 calculated using the original Hipparcos parallaxes \citep{b_per}.  As in
 Fig.~\ref{fig001}, the left and right panels are for 2MASS and DIRBE
 photometry, respectively. }
 \label{fig002}
\end{figure}

To test whether the revised Hipparcos parallaxes are superior to the
original values, we use the ability to resolve the TRGB, with consistent
results, as an indicator of precision. We reprocessed the same sample used
for Figure~\ref{fig001}, but using the original parallaxes
\citep{b_per}. Figure~\ref{fig002} shows the resulting bootstrap
distributions for 2MASS and DIRBE photometry.

We note several features. Firstly, the peaks are no longer coincident, with
the filters converging to different values. Some peaks are asymmetrical,
with long tails, and in the case of SMF96 with DIRBE data, multiple
solutions are found. None of the tip locations match the value consistently
returned using revised parallaxes. Moreover, the peaks have significantly
larger widths (0.05--0.12 mag, excluding the worst case with multiple
solutions).  Comparing with Figure~\ref{fig001} (bottom row), and noting
the change in scale, we see that only the revised parallaxes produce a
consistent result with small uncertainties. Tests using larger samples with
greater relative parallax limits produced similar results.  We conclude
that the revised Hipparcos parallaxes are superior to the originally
published values, and that this improvement is necessary for a clear
detection of the TRGB.

%%%%%%%%%%%%%%%%%%%%%%%%%%%%%%%%%%%%%%%%%
\subsection{The I-band TRGB}
%%%%%%%%%%%%%%%%%%%%%%%%%%%%%%%%%%%%%%%%%%

An $I$-band calibration of the tip is particularly important because of its
relative insensitivity to metallicity and age. Unfortunately, we are not
aware of a homogeneous source of all-sky $I$-band photometry for these very
bright stars. For example, DENIS photometry is saturated at $m_I \sim 10$
\citep{b_epc94}. Instead, we transformed 2MASS NIR magnitudes to the
$I$-band using
%an $I=f(J,H,K)$ calibration
the relation $m_I=0.0560+2.0812\;m_J+0.4074\;m_H-1.4889\;m_K$
(G. Bakos, pers. comm.), with uncertainties of $\sim$0.1--0.2\,mag. This
required a tighter constraint on $\sigma_{\pi}/\pi$ to compensate for the
larger photometric uncertainties introduced by the transformation, which
reduced the number of stars near the tip.

Subsequent edge searches yielded inconsistent results. Samples having
$\sigma_{\pi}/\pi \le 0.1$ produced very broad bootstrap distributions
around $M_I \sim -4.1$ with vague, ill-defined peaks. Samples with greater
relative parallax uncertainties produced slightly more consistent results,
with $M_I$ generally centered at $\sim -3.8 \pm 0.2$ mag, although
inspection of the corresponding LFs showed no obvious
discontinuity. Clearly an improved source of precise, all-sky $I$-band
photometry is required before at accurate calibration can be made.

%%%%%%%%%%%%%%%%%%%%%%%%%%%%%%%%%%%%%%%%%%%%%%%%%%%%%%%
\section{Conclusion}
\label{s_conc}
%%%%%%%%%%%%%%%%%%%%%%%%%%%%%%%%%%%%%%%%%%%%%%%%%%%%%%%

The TRGB is an important, widely-used tertiary distance indicator, but lacks a direct
calibration. Using revised Hipparcos parallaxes and NIR photometry from the 2MASS and DIRBE
catalogs, we have unambiguously detected the TRGB in the local solar neighborhood. Two
different ED methods have been used to quantitatively measure its location as $M_K = -6.85 \pm
0.03$, which is the first geometric calibration of the TRGB, and the first for stars with solar
metallicity. We demonstrate that the revised Hipparcos parallaxes yield consistent results with
far smaller uncertainty than the original Hipparcos parallaxes, and conclude that the revised
values are indeed superior, at least for our sample of nearby, M giants. We were unable to
detect the effect of the Lutz-Kelker bias using the TRGB location as an indicator. Calibration
of the $I$-band TRGB is not possible due to a lack of sufficiently precise photometry.

\smallskip

%%%%%%%%%%%%%%%%%%%%%%%%%%%%%%%%%%%%%%%%%%%%%%%%%%%%%%%%%%%%%%%%%%%%%%%%%%%%
%\section*{Acknowledgments}

This research has made use of the data products from the Two Micron All Sky Survey, which is a joint project of the University of Massachusetts and the
Infrared Processing and Analysis Center/California Institute of Technology, funded by the National Aeronautics and Space Administration and the
National Science Foundation. This project has been supported by the Australian Research Council.

%%%%%%%%%%%%%%%%%%%%%%%%%%%%%%%%%%%%%%%%%%%%%%%%%%%%%%%%%%%%%%%%%%%
% end of footnotesize

% APJ
%\bsp

\label{lastpage}

\end{document}